\documentclass[prx,twocolumn, superscriptaddress,amsmath,amssymb,floatfix]{revtex4-2}
\usepackage{graphicx}
\usepackage{dcolumn}
\usepackage{bm}
\usepackage{textcomp}
\usepackage{gensymb}
\usepackage{multirow}
\usepackage[none]{hyphenat}
\usepackage{epstopdf}
\usepackage{amsmath}
\usepackage[bb=dsserif]{mathalpha}
\usepackage{float}
\usepackage{tabularx}
\usepackage{color}
\usepackage{siunitx}
\usepackage{xr}
\usepackage{inputenc}
\usepackage{nicematrix}
\usepackage{blkarray}

\usepackage{amsmath} 
\usepackage{mathtools}
\usepackage{physics}
\usepackage[normalem]{ulem}
\usepackage{natbib}
\usepackage[caption=false]{subfig}
\usepackage{afterpage}
\usepackage{enumerate}
\usepackage{hyperref}
\usepackage{color,soul}
\usepackage[all]{hypcap}

\newcommand{\hc}{\hat{c}}

\begin{document}

\title{Stacked tree construction for free-fermion projected entangled pair states}

\author{Yuman He}

\author{Kangle Li}

\author{Yanbai Zhang}

\author{Hoi Chun Po}
    \email{hcpo@ust.hk}
    \affiliation{Department of Physics, Hong Kong University of Science and Technology, Clear Water Bay, Hong Kong, China}
    \affiliation{IAS Center for Quantum Technologies, Hong Kong University of Science and Technology, Clear Water Bay, Hong Kong, China}
    \affiliation{Center for Theoretical Condensed Matter Physics, Hong Kong University of Science and Technology, Clear Water Bay, Hong Kong, China}

\date{\today} 

\begin{abstract}

The tensor network representation of a state in higher dimensions, say a projected entangled-pair state (PEPS), is typically obtained indirectly through variational optimization or imaginary-time Hamiltonian evolution. 
Here, we propose a divide-and-conquer approach to directly construct a PEPS representation for free-fermion states admitting descriptions in terms of filling exponentially localized Wannier functions. 
Our approach relies on first obtaining a tree tensor network description of the state in local subregions. Next, a stacking procedure is used to combine the local trees into a PEPS. Lastly, the local tensors are compressed to obtain a more efficient description. 
We demonstrate our construction for states in one and two dimensions, including the ground state of an obstructed atomic insulator on the square lattice.
\end{abstract}


\maketitle

\section{Introduction}
Tensor network (TN) representations, like the matrix product state (MPS), provide highly efficient descriptions of quantum many-body states. 
An MPS is always disconnected when any of its bonds is cut. Physically, this implies the  virtual Hilbert space attached to any bond in the network could be given a simple interpretation in terms of the bipartite entanglement of the state {\cite{Schollw_review,Orus_review}}. 
This interpretation is valid whenever the tensor network is free of loops, and it has  two major consequences. 
First, a loop-free TN state is {\it constructible}, in the sense that one could directly construct the TN representation of any given state, up to an error threshold, by successive bipartitions of the system { \cite{vidal_classical_simulation,vidal_1d_simulation,Orus_review,Schollw_review}}. 
Second, a loop-free TN state is also {\it computable}, in that there exist canonical forms which enable the numerically exact contraction of TN diagrams arising from, for instance, the computation of physical observables {\cite{DMRG,Orus_review,Schollw_review}}.

Though powerful, the MPS ansatz is natural only for one-dimensional (1D) systems (or as a quasi-1D modeling of higher-dimensional systems). Applying the ansatz in higher dimensions through either a 1D ordering of all the sites {\cite{snake_tensor_network,other_space_filling,2d_DMRG_White}} or through its extension to a tree TN state {\cite{VerstraeteTTN,vidal2006TTN,vidal2009TTN}}, which remains loop-free, would unavoidably assign certain physically neighboring sites to far-apart nodes on the network. Such TNs are generally incompatible with the physical locality of the state and therefore cannot efficiently encode part of the short-range entanglement in the system.

The projected entangled pair state (PEPS) is another natural generalization of MPS to higher dimensions in which physical locality is retained at the cost of introducing loops \cite{Orus_review}.
As a result, a PEPS is generally neither constructible nor computable: variational optimization or imaginary-time Hamiltonian evolution is needed for finding the PEPS representation of a state, and approximations are invoked in evaluating physical observables through TN contractions  {\cite{PEPS_complexity}}.

The recent proposal on isometric TN states {\cite{2d_isometric,3d_isometric}} has provided a fruitful avenue for attacking the computability problem of PEPS.
Here, we seek to address the complementary problem concerning constructibility, namely, can one obtain a PEPS representation directly from a given state?
This question had been answered in the affirmative for the ground states of special models, like those corresponding to stabilizer codes {\cite{PEPS_topology,criticality_PEPS,TN_stabilizer}}. However, for more general problems, indirect approaches like  optimization or imaginary-time evolution remain the only tenable options so far. This is true even for free-fermion states, as is reflected in the recent bodies of work concerning fermionic Gaussian TN states {\cite{fGPEPS,fGPEPS_chiral,fGPEPS_fermi_surface,Dubail_Read_No_go,MPS_fG, quantum_impurity_fG}}.

In this work, we demonstrate the PEPS constructibility of the ground state of a free-fermion obstructed atomic insulator in two dimensions
\footnote{
By a ``free-fermion'' state, we refer to a fermionic Gaussian state with a definite particle number.
}
Our approach follows a divide-and-conquer strategy and consists of three steps, in the order of ``tree, stack, and compress.''
First, we derive the {\it tree} TN representations for the local descriptions of the state over small open disks.
Next, we {\it stack} the tree TN states to cover the full two-dimensional space. Importantly, the patches overlap and so the resulting TN takes a PEPS form. Lastly, we {\it compress} the local tensors to obtain an efficient representation. This is achieved by applying MPS techniques to the partial contractions of the TN state along one-dimensional subregions.

\begin{figure*}
    \includegraphics[width=\textwidth]{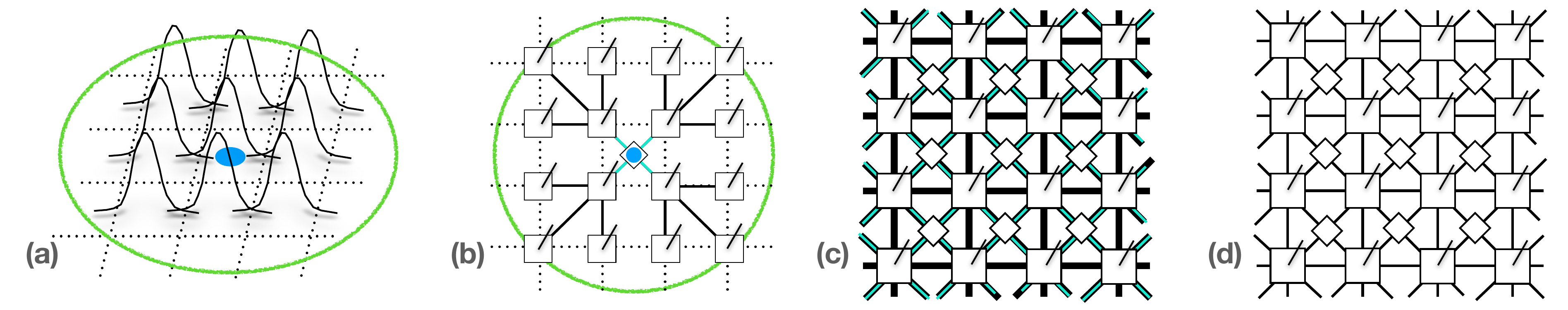}
    \caption{The overall procedure of the ``tree, stack and compress''. (a) Exponentially decaying WFs centered at the centers of plaquettes of the square lattice, while atomic sites are the vertices. (b) The tree decomposition of a single WF, where the center of the WF is indicated by a blue dot. The green circle encloses the truncated region for the WF. The dashed lines indicate the square lattice, while the solid lines are legs of the tensors. The legs pointing out-of-page are the physical legs, whereas those in-plane  are virtual legs/bonds of local tensors. 
    Blue legs connect the diamond in the center to the neighboring sites, and black legs connect among the square tensors.  (c) PEPS obtained by stacking the trees over the whole lattice using translation symmetry. Note that both blue (between square and diamond tensors) and black (among square tensors) are present along the diagonal direction.
    (d) The local tensors all over the lattice after the compression. 
    }
    \label{fig:overall}
\end{figure*}

We remark that, as a proof of principle, we consider here the ground states of translation-invariant free-fermion Hamiltonians. This allows us to shortcut some of the analysis through band-theory techniques like Wannierization. Our approach can be readily generalized to any free-fermion state admitting a (possibly approximate) localized description in terms of filled Wannier functions (WFs).
In our current formulation, however, the ``stacking'' step makes explicit use of the Gaussian nature of the local tensors; generalizing this step to an interacting state is likely nontrivial.
Nevertheless, the free-fermion TN representation we constructed could still serve as a natural starting point for constructing an interacting fermionic TN state through, for instance, the Gutzwiller projection {\cite{d_wave_Gutzwiller_HHT,Gutzwiller_HHT,Gutzwiller_chiral,parton_HHT}}.

\section{Setup}
We begin by explaining how our ``tree, stack, and compress'' steps are carried out for a free-fermion state.
{We consider a free-fermion Hamiltonian } 
$\hat{H} = \sum_{ij}h_{ij}\hat{c}_{i}^{\dagger}\hat{c}_{j}$, where $\hat{c}^{\dagger}_{i}$ and $\hat{c}_{i}$ are respectively the free-fermion creation and annihilation operators. The subscript $i$ denotes possible degrees of freedom, like physical sites, orbitals etc.
The ground state $\ket{\Psi}$ of $\hat H$, as is the case for any fermionic Gaussian state, is fully determined by its two-point correlation functions \cite{Peschel_2009}. 

{With the number conservation symmetry in our context, we only need to focus on the correlation matrix $C_{ij} = \expval{\hat{c}_{i}\hat{c}^{\dagger}_{j}}{\Psi}$. We further specialize to the case that $\hat H$ is translationally invariant and $\ket{\Psi}$ can be obtained by the filling of a full set of WFs, which corresponds to an atomic insulator.
The WFs can be viewed as a particularly suitable choice of Fourier transform of the filled Bloch states such that they become exponentially localized in the real space \cite{WF_review}. For an atomic insulator, the WFs can be chosen such that they further respect all the internal and spatial symmetries of the system.
}

\begin{figure*}
    \includegraphics[width=\textwidth]{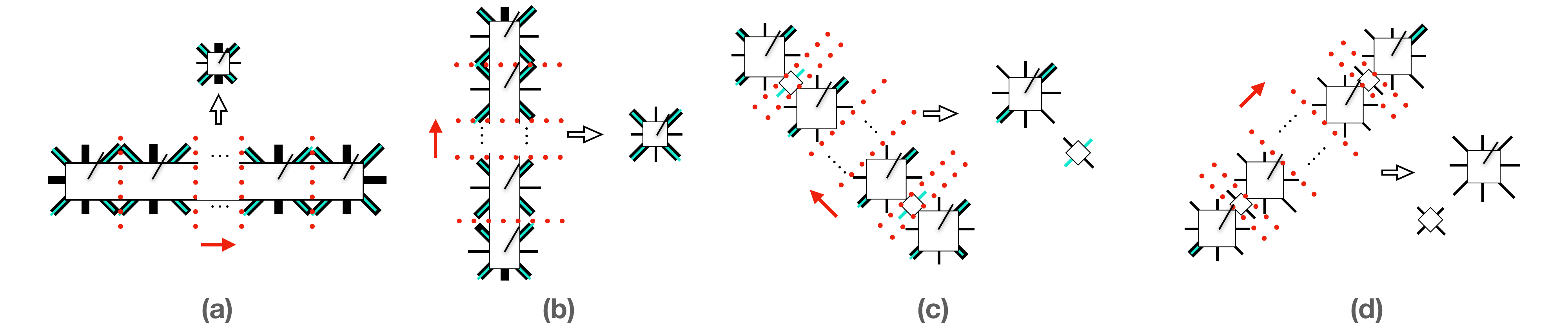}
    \caption{Compression procedure. The thickness of the legs indicates the size of the bond space. Red dashed lines indicate the successive Schmidt decomposition performed along the red arrow. The hollow arrow points to the final result of each compression step, which is the middle piece among all the decomposed local tensors. The ordering of compression direction is first along (a) horizontal direction, then (b) vertical direction and lastly along the two diagonal directions (c) and (d). }
    \label{fig:compression}
\end{figure*}

\noindent {\it Tree decomposition.}
Our first step is to obtain a tree TN description for the ground state over a small local subregion. This can be achieved by first focusing on a single WF, which represents a locally defined fermion mode that is occupied in the ground state. Given the exponential localization, the WF can be well-approximated by a truncation to a disk of some radius $r_{\rm trunc}$ which is on the order of its localization length. 
This is illustrated in Fig.\ \ref{fig:overall} (a), where the WF centered at the blue dot is picked, and the truncation is indicated by the green circle.

To obtain a TN representation of the truncated WF, we define a tree which specifies how the sites in the region are to be connected in the TN.
For instance, as demonstrated in Fig.\ \ref{fig:overall} (b), we can grow a tree with 4-fold rotation symmetry on the square lattice. 

As a tree is loop-free, we can convert the wave function into a tree TN form by successively applying Schmidt decompositions. More concretely, we view the center site of the tree, which coincides with the center of the WF, as its root. 
Note that the center need not be occupied by a physical site, and so there may not be a physical leg attached to the center (Fig.\ \ref{fig:overall}).
Any other sites can be given a height according to its distance from the root. We say two sites belong to the same level if they are equidistant from the root. Starting from the highest level, we perform Schmidt decomposition to obtain the local tensors defined on the sites in the level. Schmidt decompositions within the same level are independent. 

For free-fermion states, Schmidt decomposition can be done at the level of correlation matrices, since the reduced density matrix of a subregion is still Gaussian and so it shares the same eigenbasis with the restricted correlation matrix. More explicitly,  consider a bipartition of the system into $A$ and $B$. We can diagonalize the restricted correlation matrices of $C$ as:
\begin{equation}\label{eq:Schmidt_decomposition}
\begin{split}
    \begin{bmatrix}
        C_{AA} & C_{AB} \\
        C_{AB}^\dagger & C_{BB}
    \end{bmatrix} 
    = \begin{bmatrix}
        U_{A} & \\
         & U_{B}
    \end{bmatrix} \begin{bmatrix}
        \Omega_{AA} & \Omega_{AB}\\
        \Omega_{AB} & \Omega_{BB}
    \end{bmatrix}\begin{bmatrix}
        U_{A}^{\dagger} & \\
         & U_{B}^{\dagger}
    \end{bmatrix},
    \end{split}
\end{equation}
where $C_{AA}$, $C_{AB}$, and $C_{BB}$ are the corresponding submatrices of $C$. $\Omega_{AA}$ and $U_{A}$ are respectively the eigenvalues and diagonalizing unitary of the restricted correlation matrix $C_{AA}$; similarly for $\Omega_{BB}$ and $U_{B}$. $\Omega_{AB}$ would generally depend on the arbitrary phases in $U_A$ and $U_B$ \footnote{More generally, basis choice within degenerate subspaces of the singular values.}, but it can be brought into a real diagonal form with non-negative entries through a suitable basis choices (Appendix \ref{ap:schmidtD}). 

After Schmidt decomposition, one expects to obtain two free-fermion tensors $T_A$ and $T_B$ which can be contracted to reproduce the original state $|\Psi\rangle$. The meaning of a free-fermion tensor, however, is unclear as there will generally be multiple legs with varying number of fermion modes attached to them.
For bosonic systems, like qubits, a tensor can always be reinterpreted as a state through a mere reshaping of the legs; for fermions, care must be taken to ensure such reshaping is done in a consistent manner
\footnote{
Mapping the fermion problem to a bosonic one is one possible (and perhaps standard) way, but this would invoke unwieldy Jordan-Wigner strings and does not manifestly preserve the free-fermion nature of the present problem. While this problem is arguably mild for an MPS or a tree TN, it is much more severe for the subsequent steps in our problem concerning a 2D state. As such, a fully fermionic formulation is desired, as we discuss in Appendix \ref{sec:appendix6}.
}.
In our formulation, this is achieved by purifying all the unitary operators into fermionic Gaussian states defined on a doubled space \cite{nielsen_chuang_2010}. 
In Appendix \ref{sec:appendix6}, we show how the data contained in Eq.\ \eqref{eq:Schmidt_decomposition} can be packaged into two free-fermion states $|T_A\rangle$ and $|T_B\rangle$, which reproduce $|\Psi\rangle$ upon contraction.
This way, all the local tensors can be interpreted as free-fermion states. As such, in the following we use ``local tensors'' interchangeably with the correlation matrix of its corresponding free-fermion state.

Upon performing all the contractions of the local tensors, as represented by the edges on the tree in Fig.\ \ref{fig:overall}(b), we reconstruct the single-particle state given by filling the original (truncated) WF in the current local subregion.

\noindent {\it Stacking.} To reconstruct the full state $\ket{\Psi}$, we would need to combine the locally defined tree TN states obtained from the individual WFs. Intuitively, we simply need to consider the collection of all the tree TN states, which in our context are related to each other through translation symmetry, and show that these states can be recombined into a single PEPS (Fig.\ \ref{fig:overall} (c)). This step, referred to as the ``stacking'' procedure, can be achieved as follows.
Suppose the local tensor $C^{i,\alpha}$ at site $\alpha$ decomposed from the $i^{th}$ tree is represented as:
\begin{equation} 
C^{i,\alpha} = \left[\begin{array}{c|c}
        \Gamma^{i,\alpha}_{\rm pp} & \Gamma^{i,\alpha}_{\rm pv} \\
        \hline
        {\Gamma^{i,\alpha}_{\rm pv}}^{\dagger} & \Gamma^{i,\alpha}_{\rm vv}\\
        \end{array}
        \right]
\end{equation}
where the correlation matrix is organized with respect to the physical legs and bonds. Here, $i$ indexes the set of truncated WFs which have support on the site $\alpha$, and the subscripts ${\rm p}$ vs. ${\rm v}$ indicate whether the fermion modes are associated with the physical or the virtual legs.

The stacking of the local tensors $C^{i,\alpha}$ for the site $\alpha$ could be expressed as:
\begin{align}
\tilde{C}^{\alpha} &= \left[ 
        \begin{array}{c|cccc}
        \sum_{i=1}^m \Gamma^{i,\alpha}_{\rm pp} & \Gamma^{1, \alpha}_{\rm pv} & \Gamma^{2, \alpha}_{\rm pv} & \cdots & \Gamma^{m, \alpha}_{\rm pv}\\
        \hline
        {\Gamma^{1, \alpha}_{\rm pv}}^\dagger & \Gamma^{1,\alpha}_{\rm vv} & 0 & \cdots  & 0\\
        {\Gamma^{2, \alpha}_{\rm pv}}^\dagger &  0 & \Gamma^{2,\alpha}_{\rm vv} & \ddots & \vdots\\
        \vdots & \vdots & \ddots & \ddots& 0\\
        {\Gamma^{m, \alpha}_{\rm pv}}^\dagger & 0 & \cdots & 0 & \Gamma^{m,\alpha}_{\rm vv}
        \end{array}
    \right]
\end{align} 
where we suppose that there are $m$ trees that contribute to the site $\alpha$. 

Here, the virtual spaces from different trees are independent and so the corresponding parts of their correlation matrices are simply combined as a direct sum. However, all the trees share the same physical Hilbert space at site $\alpha$ and so their contributions add up. As defined, $\Tilde{C}^\alpha$ is not a proper correlation matrix in general, as the summing procedure defined above does not correspond to any well-defined operations on the Hilbert spaces concerned. Nevertheless, a proper correlation matrix $C^{\alpha}$ can be obtained by the deformation procedure described in Appendix \ref{AP:deformation}. Intuitively, the failure of $\Tilde{C}^\alpha$ to be a proper correlation matrix stems from the fact that the restrictions of the different trees to the physical Hilbert space of site $\alpha$ lead to modes which are not orthogonal to each other. The deformation process can then be simply interpreted as a suitable orthonormalization step.

We thus obtain a free-fermion PEPS defined by the collection of deformed local tensors. Upon contracting all the virtual legs, we obtain an approximation of the ground state $\ket{\Psi}$.

\noindent {\it Compression.}
The approximate PEPS representation obtained from stacking, however, is far from optimal. In combining the individual trees, we treated their virtual Hilbert spaces as independent. This leads to a superficially high bond dimension which grows as we increase the truncation radius $r_{\rm trunc}$ used in approximating the WFs (Appendix \ref{AP:WF}). As a last step, therefore, we perform a compression of the local tensors. 

The idea is that we could first contract the local tensors in one direction to form a free-fermion state defined on an open 1D chain. We can then perform another MPS decomposition of the state 
while retaining only the most significant virtual modes, which correspond to a truncation to the bond dimension.
More concretely, the virtual modes are retainiend according to their contribution to the von Neumann entanglement entropy: $S = -(\sum_{i}\omega_{i}\ln\omega_{i} + (1-\omega_{i})\ln(1-\omega_{i}))$ \cite{Peschel_2009,Latorre_2009,PhysRevLett.90.227902,10.5555/2011572.2011576,Fishman_compression_fG}, where $\omega_{i}$ corresponds to $i^{th}$ diagonal entry for matrix $\Omega_{AA}$ or $\Omega_{BB}$ in equation \ref{eq:Schmidt_decomposition}. To reduce the bond dimension, we drop virtual modes that contribute the least to the entanglement entropy, i.e., we drop the mode $i$ if $\omega_i < \epsilon$ or $\omega_i > 1-\epsilon$ for a prescribed small threshold $\epsilon$.
Physically, these dropped virtual modes correspond to degrees of freedom that are well-localized within one side of the entanglement cut, i.e., they do not mediate entanglement across the cut and can therefore be dropped.

After successive Schmidt decomposition, local tensors deeply embedded in a long enough 1D chain should regain bulk properties. Therefore, we choose the local tensor in the middle as the updated $C^{\alpha}$. Repeating the above process along all possible directions for multiple times, the final fully-compressed $C^{\alpha}$ is obtained, see Fig.\ \ref{fig:compression}.

\begin{figure}[!ht]
	\centering
	\includegraphics[width=1.0\linewidth]{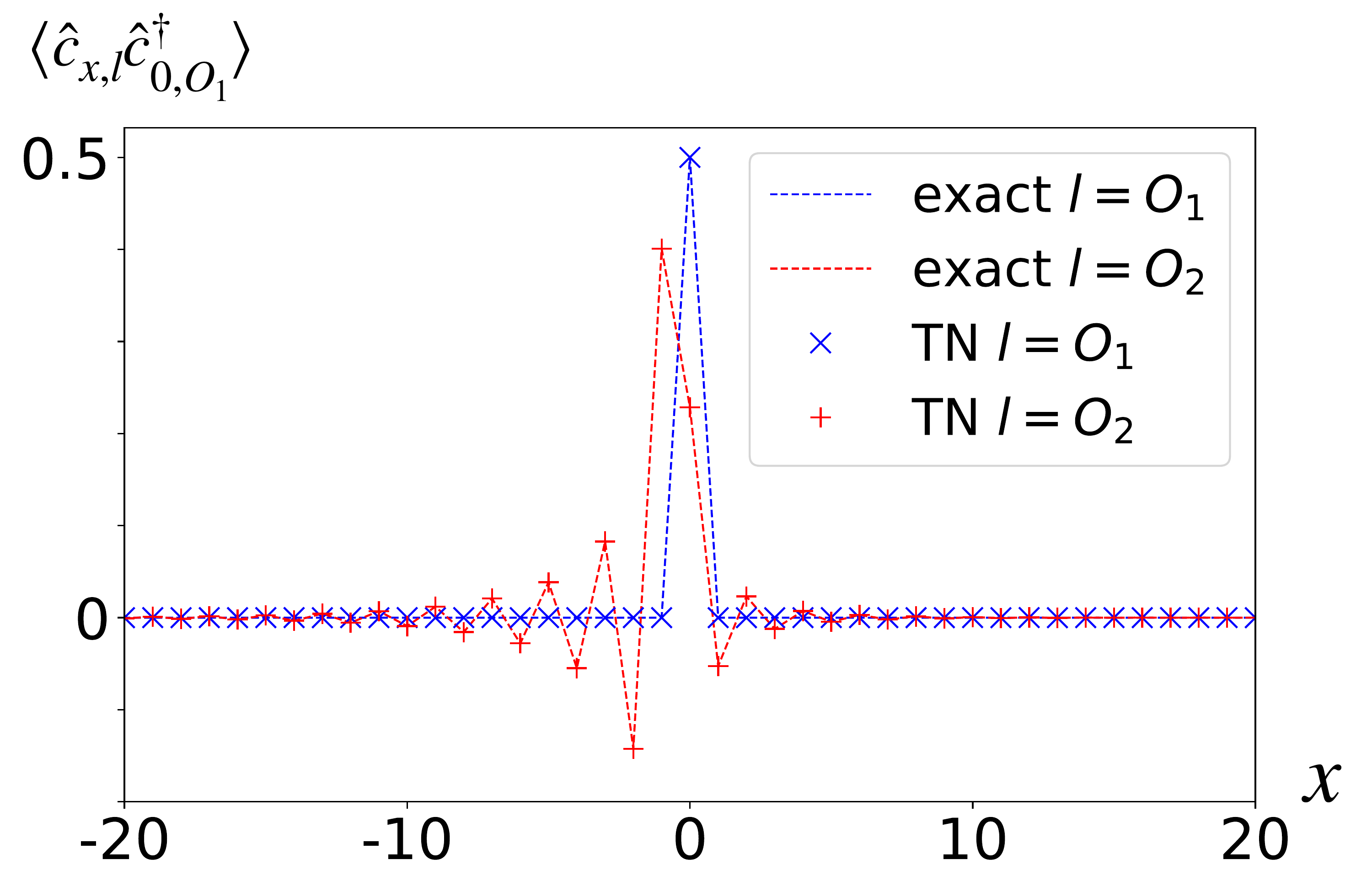}
	\caption{Correlation function $\langle\hat{c}_{x,l}\hat{c}^\dagger_{0, O_{1}}\rangle$ for a 100-site SSH model. Only data inside the range $(-20,20)$ is 
 {shown as the values are practically zero outside of this range}. The blue and red colors represent the two orbitals $O_{1}$ and $O_{2}$ respectively. The dashed lines represent exact values while the markers represent the compressed ones obtained from the constructed TN.}
	\label{fig:SSH_corr_comparison}
\end{figure}

\begin{table*}[t]
\centering
\begin{tabular}{ |p{3.5cm}||p{2cm}|p{2cm}|p{3cm}|p{3cm}| }
 \hline
 & \multicolumn{2}{|c|}{1D SSH} & \multicolumn{2}{|c|}{2D OAI}\\
 \hline
 & original & compressed & original & compressed\\
 \hline
 System Size & 100 & 100 & 50 $\times$ 50 & 100 $\times$ 100\\
 \hline
$r_{\rm trunc}$& \multicolumn{2}{|c|}{16}  & \multicolumn{2}{|c|}{3}  \\
 \hline
$b$ & $b=31$   & $b=7$ &  $b_{\rm v, h}=12$; $b_{\rm d}=3$&$b_{\rm v, h}=5$; $b_{\rm d}=3$ \\

\hline
$\max_{i,j}(|C_{TN} - C_{exact}|_{ij})$  & 6.56 $\times 10^{-4}$   &  1.45 $\times 10^{-4}$ &  3.16 $\times 10^{-3}$ & 4.44 $\times 10^{-3}$\\
\hline
$(e_{TN}-e_{exact})/\delta E_{gap}$&   3.90 $\times 10^{-4} \%$  & 6.36 $\times 10^{-4} \%$   & 0.17\% & 0.32\%\\
 \hline
\end{tabular}
\caption{Results for 1D SSH model and 2D OAI model before and after compression. Note that a larger system size is used for the 2D compressed result.
The compression thresholds for the 1D and 2D models are respectively $\epsilon=10^{-6}$ and $10^{-4}$.
}
\label{table:overall}
\end{table*}

\section{Examples} We now move on to demonstrating our construction to two obstructed atomic insulator (OAI) systems as a proof of principle. OAI is a special class of atomic insulators for which the centers of the WFs cannot be chosen to coincide with any atomic sites in the system \cite{fillingenforced_OAI}. 

The 1D Su–Schrieffer–Heeger (SSH) model, one of the most well-known model for a topological insulator, can also be viewed as an OAI if only inversion but not chiral symmetry is retained.

 The Hamiltonian for the SSH model is $\hat{H}=\sum_{x}(t-s) \hat{c}^{\dagger}_{O_{1}, x}\hat c_{O_{2}, x} + (t+s) \hat c^{\dagger}_{O_{2},x}\hat c_{O_{1}, x+1} + h.c.$, where $t$ is the uniform hopping parameter, and $s$ is a staggering between intra- and inter-cell hoppings. $O_{1}$ and $O_{2}$ are two different orbitals in a unit cell. In the numerics, we choose $t=-1$, $s=0.1$.

As a second example, we construct an OAI model on the 2D square lattice protected by the four-fold rotation symmetry $C_4$. 
We assign three fermion modes, corresponding respectively to $s$, $d_{x^{2} - y^{2}}$ and $p_{x} + ip_{y}$ atomic orbitals, to each of the sites.
Our model is constructed by lowering the energy of a set of non-orthogonal ``quasi-orbitals'' $\hat f^\dagger$ which transform differently from all of the atomic orbitals in the system {\cite{po2017symmetry, topological_quantum_chemistry, fillingenforced_OAI}}. This leads to a band insulator for which the WFs of the filled band are equivalent, symmetry-wise, to the quasi-orbitals we started from. More concretely, the Hamiltonian is $\hat H =-\sum_{\vec R}\eta \hat f^{\dagger}_{\vec R}\hat f_{\vec R}$, where the mode $\hat f^\dagger_{\vec R}$ is localized to the center of the plaquette in unit cell $\vec R$, and transforms trivially under the $C_4$ rotation symmetry. An atomic insulator obtained by filling $s$-like WFs localized to the centers of the plaquettes is topologically distant from the innate atomic insulators in the Hilbert space, and so this Hamiltonian serves as an OAI model (Appendix \ref{ap:2d_OAI}). In the numerical calculation, we set $\eta=2$.

\begin{figure}[!ht]
\centering
	\includegraphics[width=1.0\linewidth]{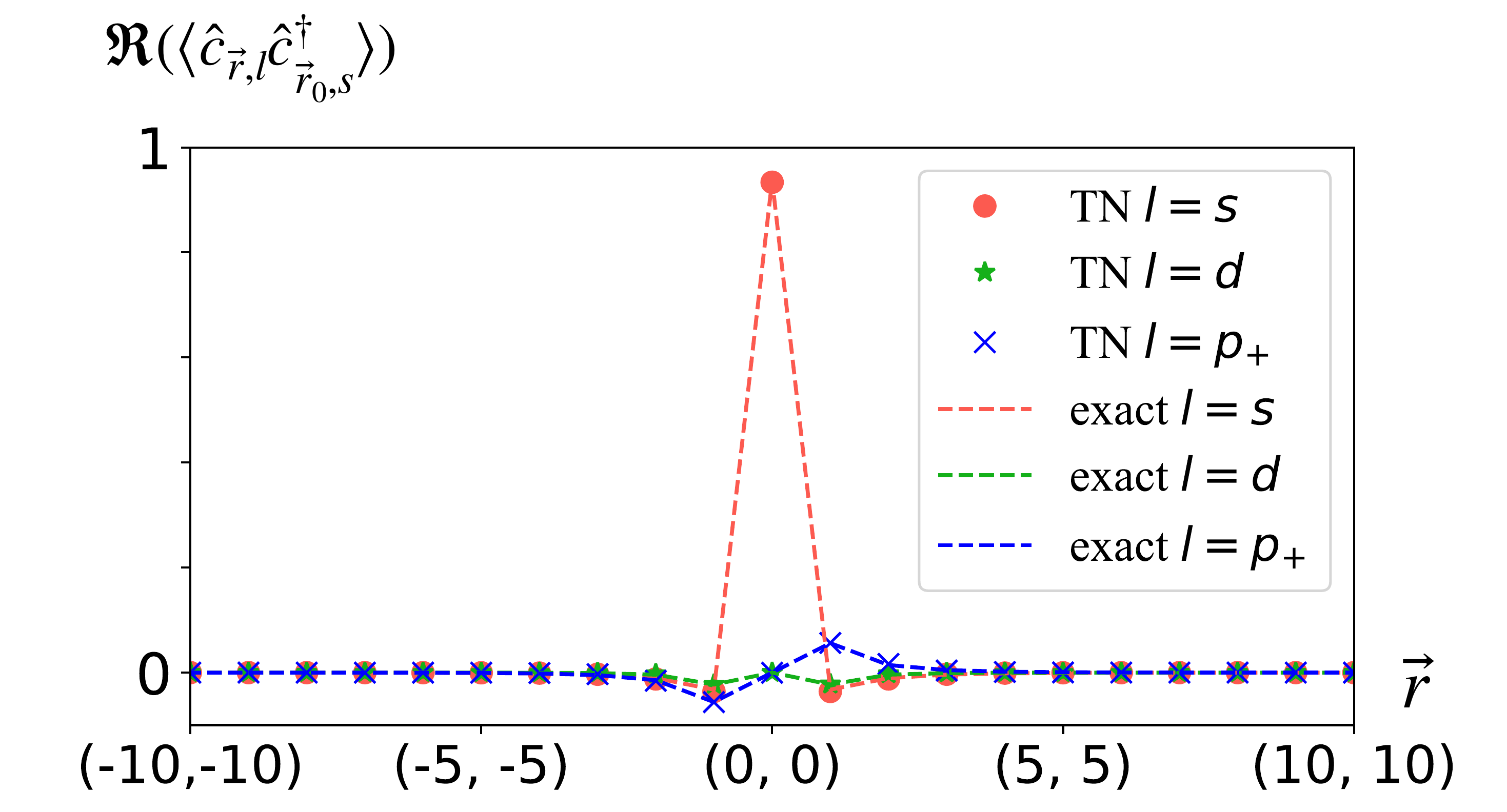}
	\caption{Real part of correlation function $\mathfrak{R} (\langle\hat{c}_{\Vec{r},l}\hat{c}^\dagger_{\vec r_{0}, s}\rangle)$ on a 100$\times$100 square lattice, where $\Vec{r}_{0} = (0, 0)$. Data for $\Vec{r}$ along direction $[11]$ within the range $(-10,10)$ and $(10,10)$ is demonstrated. $\mathfrak{R} (\langle\hat{c}_{\Vec{r},l}\hat{c}^\dagger_{\vec r_{0}, s}\rangle)$ is 0 for $\Vec{r}$ outside the region. The red, green and blue colors represent the $s$, $d_{x^{2}-y^{2}}$ and $p_+=p_{x} + ip_{y}$ orbitals respectively. The dashed lines are for exact values and the markers show the reconstructed results obtained from our TN.}
	\label{fig:OAI_corr_comparison_100}
\end{figure}

The main results are tabulated in Table \ref{table:overall}. The number of fermion modes attached to a virtual leg is denoted by $b$. For the 2D model (Fig.\ \ref{fig:overall}), we distinguish the vertical/horizontal bond $b_{\rm v,h}$ and the diagonal bond $b_{\rm d}$ for each square tensor. Before and after compression, we compare the difference between the correlation reproduced from PEPS $C_{\rm TN}$ and the exact through the entry-wise maximum difference $\max_{i,j}(|C_{\rm TN} - C_{\rm exact}|_{ij})$. Small values of this difference, on the order of $10^{-3}$, were found, and so the exact and TN results are indiscernible in Figs.\ \ref{fig:SSH_corr_comparison} and \ref{fig:OAI_corr_comparison_100}. We also compare the relative difference of ground state energy density, defined as $(e_{\rm TN}-e_{\rm exact})/\delta E_{\rm gap}$ where $e_{\rm TN}$ and $e_{\rm exact}$ are ground state energy density per unit cell obtained from TN representation and exact calculations respectively. $\delta E_{\rm gap}$ is the minimal gap over the first Brillouin zone.

\section{Discussion}
In this work, we present a general scheme for constructing PEPS for 
{free-fermion states arising from the filling of exponentially localized WFs}. As a proof of principle, we demonstrate our approach for models in one and two dimensions. 

Although translation invariance was used to simplify the computation, our approach can be generalized to a strictly real-space formulation for more general systems with incommensurate order or disorder. 
Interaction effects could also be incorporated by combining the free-fermion TN state with, for instance, Gutzwiller projectors \cite{Gutzwiller_HHT, d_wave_Gutzwiller_HHT,Gutzwiller_chiral}.
In closing, we remark that, in the restricted context of free-fermion states, there might be tantalizing connections between our construction and a tensor-network-based solution to the quantum marginal problem {\cite{PRXQuantum.2.040331,PRXQuantum.2.040331_41,PRXQuantum.2.040331_42,PRXQuantum.2.040331_43,PRXQuantum.2.040331_44,PhysRevX_quantum_marginal_problem,kim2021entropy_quantum_marginal_problem}}: data confined to small local subregions are first handled by tree TN states, which are then patched into the full pure state through the stacking step.
It is an interesting question to consider how our approach might be generalized to attack the corresponding many-body problem.

\section*{Acknowledgements} \label{sec:acknowledgements}
    This work is supported by the Ministry of Science and Technology, China through MOST22SC01 and the Hong Kong Research Grant Council through ECS 26308021.

\bibliography{bib.bib}

\appendix

\section{Details of The Models} \label{sec:appendix1}

\subsection{The minimal 3-band Hamiltonian}\label{ap:2d_OAI}

We construct a minimal 3-band ``flat-band'' model for an OAI on the square lattice in the main text. The word ``flat-band'' indicates that we only consider the dispersion of the filled band, which is obtained by filling s-like WFs, leaving the rest of bands dispersionless. Given the space group $P_{4}$, we derive a model with the band representation of the ground state induced from the Wyckoff position $1b: (\frac{1}{2},\frac{1}{2})$. Meanwhile, atomic sites of the square lattice are at Wyckoff position $1a:(0,0)$. This is required by the definition of OAI, whose band representations should be induced from the unoccupied Wyckoff positions \cite{fillingenforced_OAI}. Then by comparing the irreducible representations of the little groups at high-symmetry points in momentum space, we only need three different kinds of atomic orbitals from Wyckoff position $1a$, namely $s$, $d_{x^{2} - y^{2}}$ and $p_{x} + ip_{y}$. 

\begin{figure}[!ht]
	\centering
	\includegraphics[width=0.4\linewidth]{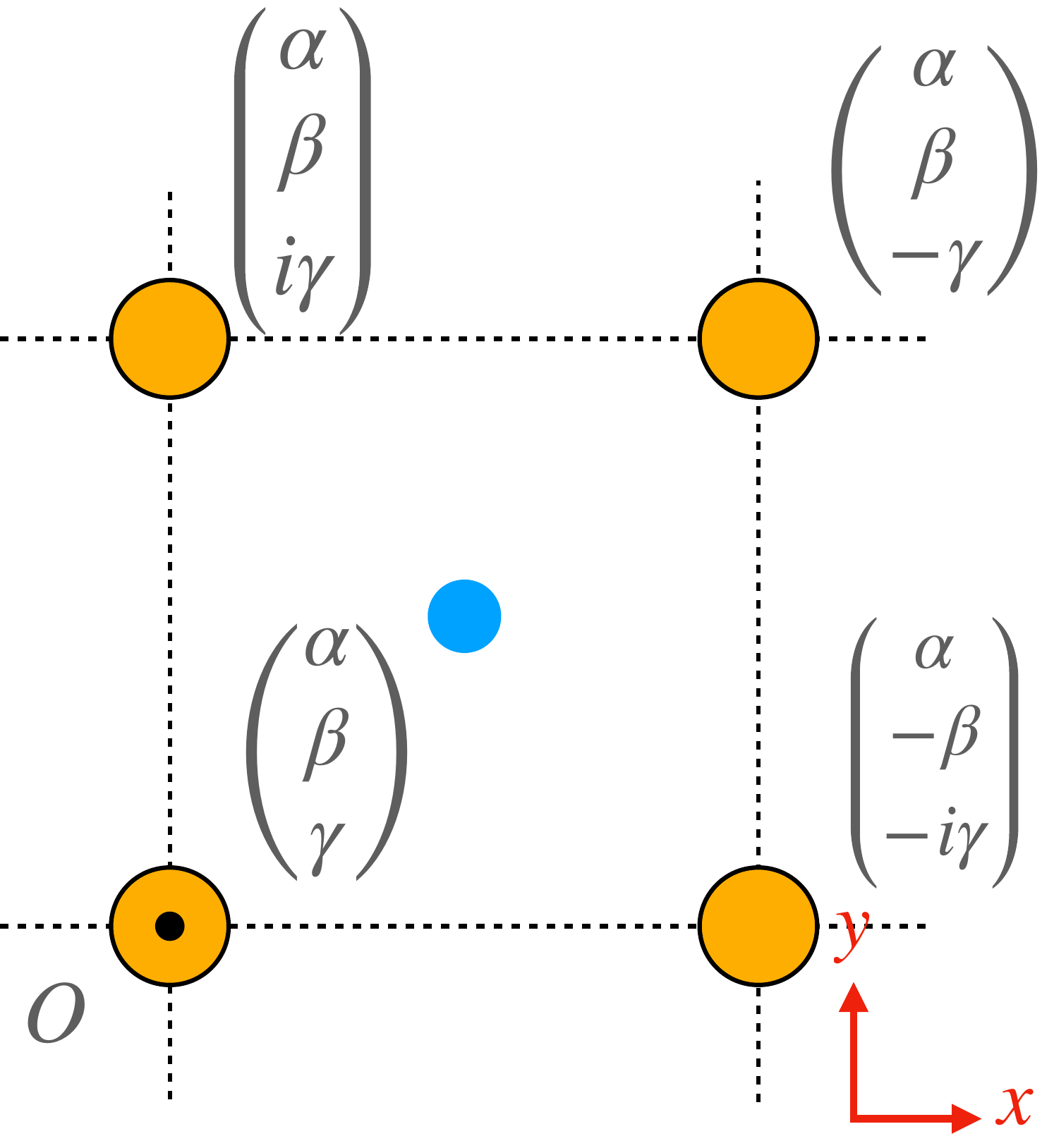}
	\caption{One unit cell of the square lattice. The balck dot $O$ is the origin, and the coordinate system is built according to the directions of the red arrows. Yellow dots at vertices are the atoms in the square lattice, which are at the Wyckoff position $1a:(0, 0)$. The blue dot is the center of the WF for the current unit cell and is at the Wyckoff position $1b: (\frac{1}{2}, \frac{1}{2})$. The vectors are local wavefunctions $w^{\dagger}_{\Vec{r}}$ for each atom site $\Vec{r}$ with the basis $s$, $d_{x^{2}-y^{2}}$ and $p_{x}+ip_{y}$ respectively. i.e. for the atom at $\Vec{r}=(0,0)$, the wavefunction $
 \hat w^{\dagger}_{\Vec{r}}$ is $\hat w^{\dagger}_{\Vec{r}}=\alpha \hat c^{\dagger}_{s,\Vec{r}} + \beta \hc^{\dagger}_{d_{x^{2}-y^{2}},\Vec{r}} + \gamma \hc^{\dagger}_{p_{x}+ip_{y}, \Vec{r}}$.
    }
	\label{fig:OAI_H}
\end{figure}

To obtain the Hamiltonian, we could start from constructing a quasi-orbital $\hat{f}^{\dagger}_{s,\Vec{R}_{o}}$ centered at point $\Vec{R}_{o}=(\frac{1}{2}, \frac{1}{2})$, consisting of atomic orbitals from the nearest atom sites, as shown in Fig. \ref{fig:OAI_H}. 
Column vectors $w_{\Vec{r}}^{\dagger}$ indicate the local hybridization for each quasi-orbital at atomic site $\Vec{r}$. The basis of column vectors is $\hc_{s,\Vec{r}}^{\dagger}$, $\hc_{d_{x^{2}-y^{2}},\Vec{r}}^{\dagger}$ and $\hc_{p_{x} + ip_{y}, \Vec{r}}^{\dagger}$ respectively from top to the bottom. The expression of the quasi-orbital using column vectors is $\hat{f}^{\dagger}_{s, \Vec{R}_{o}} = \frac{1}{2\sqrt{|\alpha|^{2} + |\beta|^{2} + |\gamma|^{2}}}(\hat w^{\dagger}_{\Vec{r}_{1}} + \hat w^{\dagger}_{\Vec{r}_{2}} + \hat w^{\dagger}_{\Vec{r}_{3}} + \hat w^{\dagger}_{\Vec{r}_{4}})$ for the unit cell at the origin, where $\Vec{r}_{1} = (0,0)$, $\Vec{r}_{2} = (1,0)$, $\Vec{r}_{3} = (0,1)$ and $\Vec{r}_{4} = (1,1)$.

\begin{figure}[!ht]
	\centering
	\includegraphics[width=1.0\linewidth]{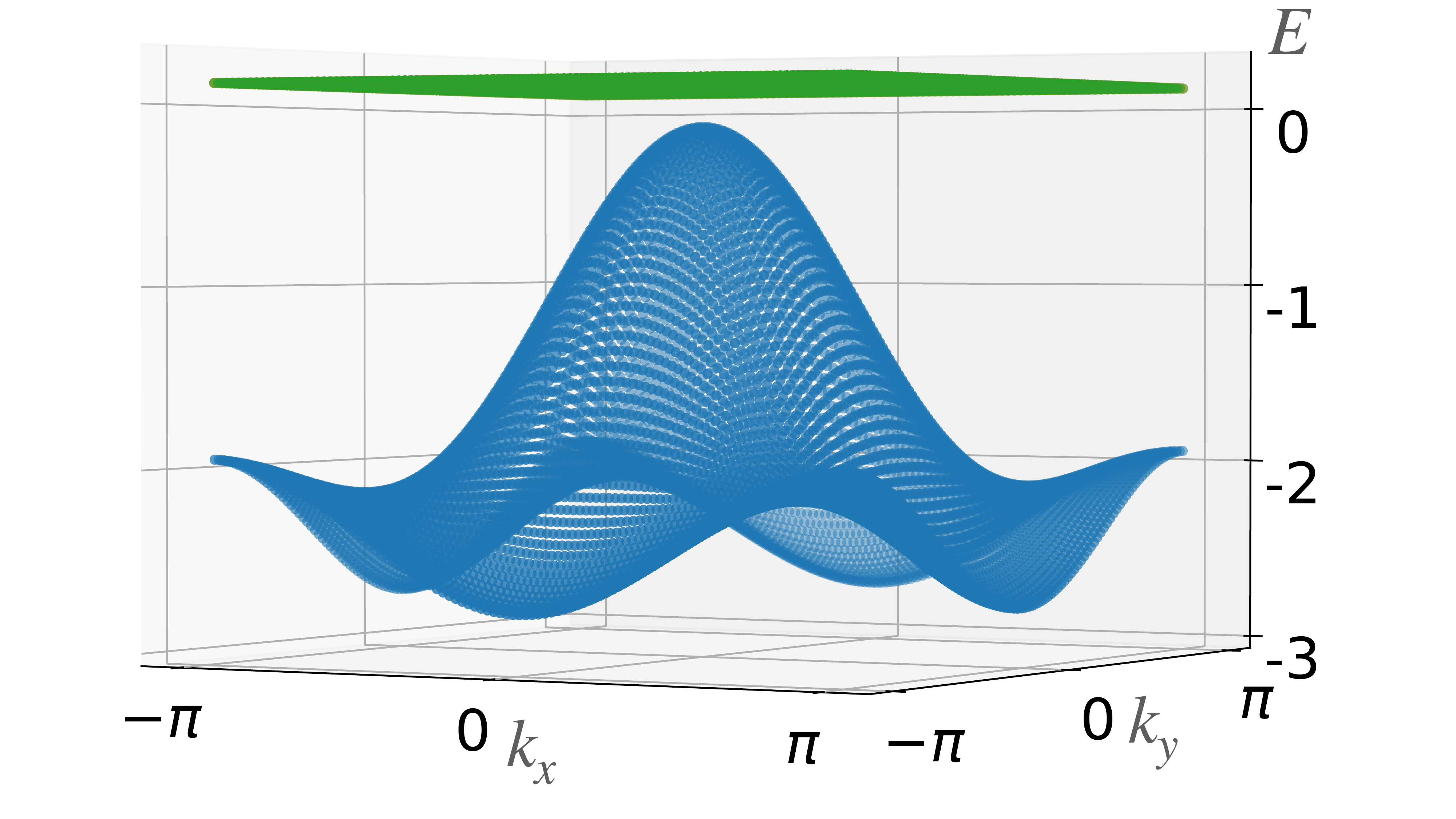}
	\caption{Band structure of the 2D OAI model on a $100 \times 100$ lattice. The smallest band gap over the first Brillouin zone is 0.2286. Bands are shifted according to the Fermi energy -0.1143.}
	\label{fig:bs}
\end{figure}

With the quasi-orbitals all over the lattice, we construct the minimal Hamiltonian:
\begin{equation}
   \hat H =-\sum_{\Vec{R}}\eta \hat{f}^{\dagger}_{s,\Vec{R}}\hat{f}_{s,\Vec{R}}, \label{eq:OAI_H}
\end{equation} 
where we summing over all unit cells $\Vec{R}$. In the numerical calculation, we have $\eta=2$, $\alpha=1$, $\beta=3$ and $\gamma=5$. The band structure across the first Brillouin zone is shown in Figure \ref{fig:bs}, where it is evident that a significant gap exists throughout the zone, providing the evidence that the model represents an insulator.

\subsection{The Wannier Function}\label{AP:WF}

We could solve the model in eq. \ref{eq:OAI_H} for eigenvalue spectrum and the Bloch states $\ket{\psi_{l}(\Vec k)}$, by first applying Fourier transform and then eigenvalue decomposition. However, a direct Fourier transform to $\ket{\psi_{l}(\Vec k)}$ cannot guarantee a set of well-localized WFs $\ket{\Psi_{l}(\Vec r)}$ in the real space, as WFs are well localized only when the Bloch functions possess a smooth gauge \cite{WF_review}. We could smoothen the phase associated to $\ket{\psi_{l}(\Vec k)}$ during Fourier transform by using trial functions $\ket{\phi_{\Vec{R}}(\Vec r)}$ to perform a projection.  $\ket{\phi_{\Vec{R}}(\Vec r)}$ is obtained through a rough guess of the distribution of $\ket{\Psi_{l}(\Vec r)}$. For instance, we choose the quasi-orbital $\ket{\phi_{\Vec{R}}(\Vec r)} = \hat{f}^{\dagger}_{s,\Vec{R}}\ket{0}$ in Appendix \ref{eq:OAI_H} as the trial function for the 2D OAI model. We first Fourier transform $\ket{\phi_{\Vec{R}}(\Vec r)}$ into $\ket{\phi_{l}(\Vec k)}$, then the Bloch states with smooth gauge $\ket{\Tilde{\psi}_{l}(\Vec k)}$ could be attained:
\begin{equation}\label{eq:WF_Projection}
    \ket{\Tilde{\psi}_{l}(\Vec k)} = \sum_{l'}\ket{\psi_{l'}(\Vec k)}\frac{\bra{\psi_{l'}(\Vec k)}\ket{\phi_{l}(\Vec k)}}{|\bra{\psi_{l'}(\Vec k)}\ket{\phi_{l}(\Vec k)}|}.
\end{equation}
Thus, WFs are:
\begin{equation}
    \ket{\Psi_{l}(\Vec r)} = \Omega\sum_{\Vec k}\ket{\Tilde{\psi}_{l}(\Vec k)},
\end{equation}
where $\Omega$ is a normalization factor. 

However, it is worth noting a technical subtlety that the projection fails when the overlap $\bra{\psi_{l'}(\Vec k)}\ket{\phi_{l}(\Vec k)}$ between the trial function and Bloch state is 0. Additionally, since we use exponentially-decaying non-compact WFs, we set the threshold to 0.1 during the projection. If $min(\bra{\psi_{l'}(\Vec k)}\ket{\phi_{l}(\Vec k)}) > 0.1$, We assert that the symmetric WFs with exponential tails have been successfully obtained.

\begin{figure}[!ht]
	\centering
	\includegraphics[width=1.0\linewidth]{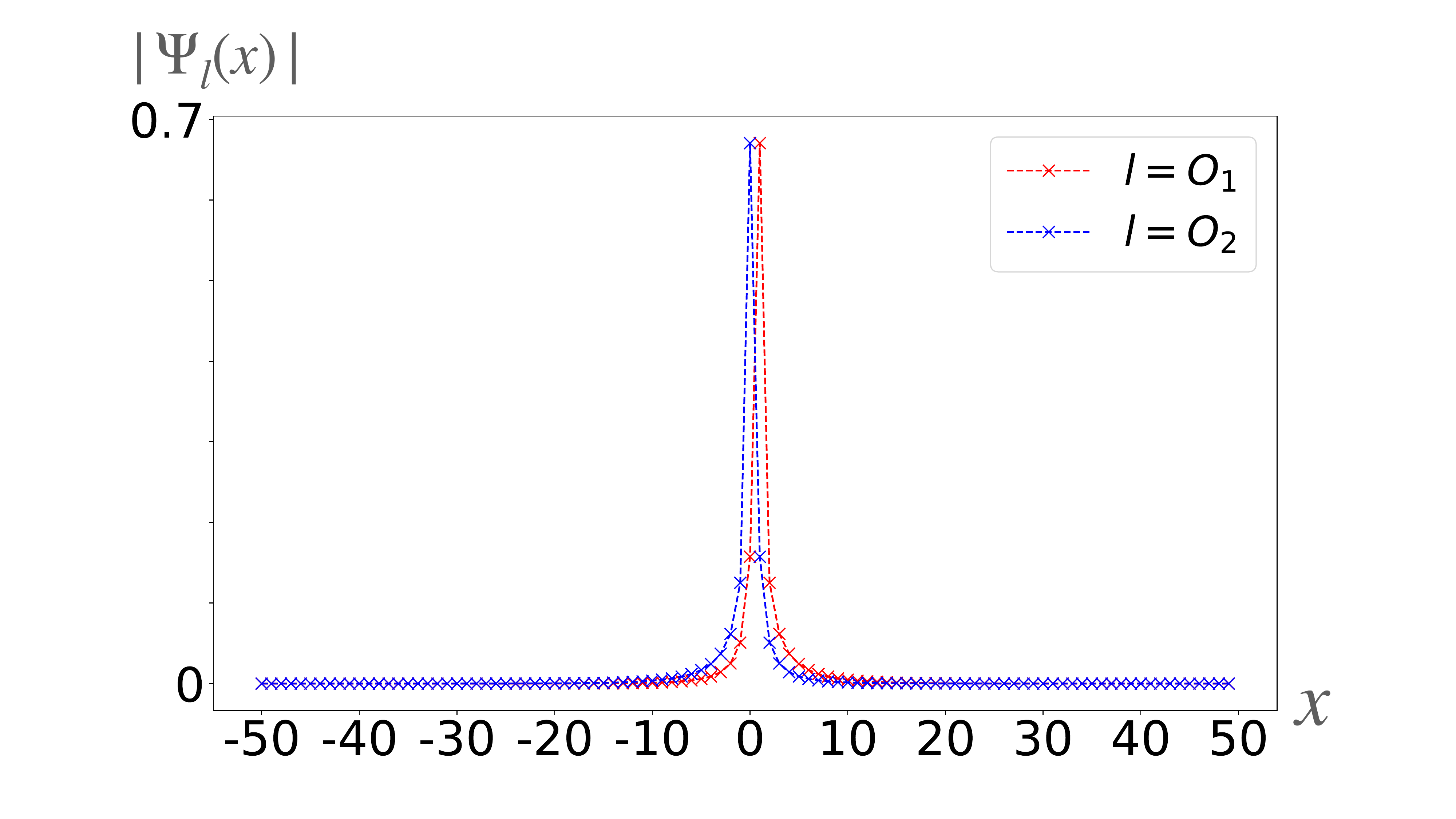}
	\caption{WF spread $|\Psi_{l}(x)|$ for the 100-site SSH model. $l$ represents the two orbitals $O_{1}$ and $O_{2}$ respectively.}
	\label{fig:SSH_WF_Spread}
\end{figure}

\begin{figure}[!ht]
	\centering
	\includegraphics[width=1.0\linewidth]{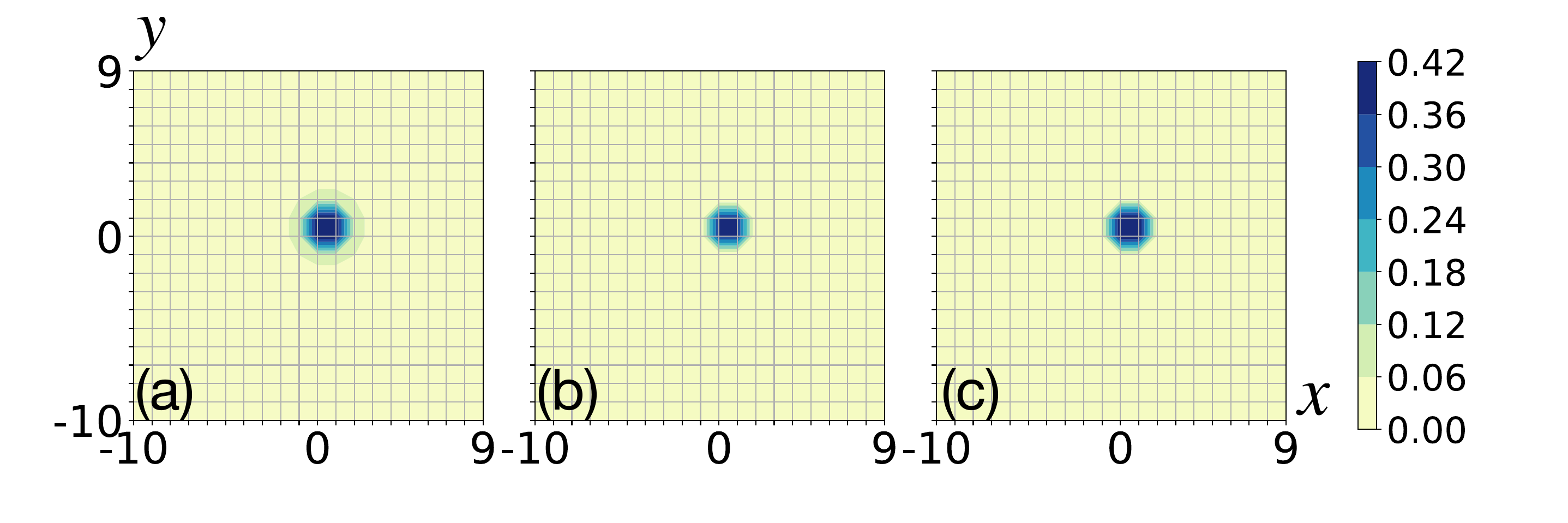}
	\caption{ WF spread $|\Psi_{l}(\Vec r)|$ on a 20 × 20 square lattice, where $\Vec r = (x,y)$. (a), (b) and (c) are for $l=s$, $l=d_{x^{2} - y^{2}}$ and $l=p_{x} + ip_{y}$ orbitals respectively.}
	\label{fig:OAI_WF_spread}
\end{figure}

We plot one WF centered at 0 and $(\frac{1}{2}, \frac{1}{2})$ for 1D and 2D model respectively, as shown in Figure \ref{fig:SSH_WF_Spread} and \ref{fig:OAI_WF_spread}. It is observed that WFs for each orbital undergo rapid decay while retaining symmetry. 

Due to the exponential localization of WFs, it is feasible to truncate them within a reasonable radius. However, truncating WFs necessitates a tradeoff between bond dimension and accuracy in the construction of PEPS. Generally speaking, larger truncation radii lead to larger bond spaces and more accurate results, as evidenced in Tables \ref{table:SSH} and \ref{table:OAI}.

\begin{table}[h!]
\centering
\begin{tabular}{ |p{3cm}||p{1cm}|p{1cm}|p{1cm}|  }
 \hline
 \multicolumn{4}{|c|}{Truncation Error List} \\
 \hline
Truncation Radius $r_{\rm trunc}$& 20 & 16 & 12\\
 \hline
Bond Dimension $b$ & 39   & 31 &  23\\
\hline
Entry-wise maximum difference $\max_{i,j}(|C_{\rm TN} - C_{\rm exact}|_{ij})$  & 2.49 $\times 10^{-4}$   & 6.56 $\times 10^{-4}$ &  1.81 $\times 10^{-3}$\\
\hline
Relative Ground State Energy Density Difference $(e_{\rm TN}-e_{\rm exact})/\delta E_{\rm gap}$ &   5.63 $\times 10^{-7}$  & 3.90 $\times 10^{-6}$   & 2.97 $\times 10^{-5}$\\
 \hline
\end{tabular}
\caption{Truncation error list of PEPS for 100-site SSH model.}
\label{table:SSH}
\end{table}

\begin{table}[h!]
\centering
\begin{tabular}{ |p{3cm}||p{1cm}|p{1cm}|p{1cm}|  }
 \hline
 \multicolumn{4}{|c|}{Truncation Error List} \\
 \hline
Truncation Radius $r_{\rm trun}$& 7 & 5 & 3\\
 \hline
Horizontal/Vertical Bond Dimension $b_{h}$ & 68   & 32  & 12\\
\hline
Diagonal Bond Dimension $b_{d}$ & 9  & 7 & 3\\
\hline
Entry-wise maximum difference $\max_{i,j}(|C_{\rm TN} - C_{\rm exact}|_{ij})$  & 1.20 $\times 10^{-4}$    & 5.00 $\times 10^{-4}$ & 3.16 $\times 10^{-3}$\\
\hline
Relative Ground State Energy Density Difference $(e_{\rm TN}-e_{\rm exact})/\delta E_{\rm gap}$&   9.19 $\times 10^{-6}$  & 1.29 $\times 10^{-4}$ & 1.73 $\times 10^{-3}$\\
 \hline
\end{tabular}
\caption{Truncation error list of PEPS for 2D OAI on a $20 \times 20$ square lattice.}
\label{table:OAI}
\end{table}

\section{Operations on fG States with Diagrams}
\label{sec:appendix6}

Fermionic Gaussian (fG) states could be expressed using density operators in the thermal form:
\begin{equation}\label{eq:fG_state}
\hat{\rho} = \frac{e^{-\hat{H}}}{Z},\\
\end{equation}
where $\hat{H}$ is a fermionic quadratic Hamiltonian and $Z = \Tr [e^{-\hat{H}}]$. The properties of fG states are totally defined by the correlation functions. Since Wick's theorem holds for fG states, it suffices to only take the two-point correlation functions into account. Furthermore, we impose the number-conserving condition, thus $\hat{H}$ becomes a free-fermion Hamiltonian: $\hat{H} = \sum_{i,j}h_{ij}\hat{c}_{i}^{\dagger}\hat{c}_{j}$, with $\hat{c}_{i}^{\dagger}$ ($\hat{c}_{i}$) the fermion creation (annihilation) operator. Meanwhile either $C'_{ij} = \langle\hat{c}^{\dagger}_{i}\hat{c}_{j}\rangle$ or $C_{ij}=\langle\hat{c}_{i}\hat{c}^{\dagger}_{j}\rangle$ is sufficient in describing fG states, as the two correlation functions are related through $C'_{ij} = \mathbb{1} - C_{ij}$. For simplicity, when we refer to the correlation functions later in the text, we explicitly refer to $C_{ij}$. For a pure state $\ket{\Psi}$, the correlation function could be expressed as $C_{ij} = \expval{\hat{c}_{i}\hat{c}^{\dagger}_{j}}{\Psi}$, while more generically, the correlation function with respect to the density matrix $\hat{\rho}$ of the state is $C_{ij} = \Tr [\hat{c}_{i}\hat{c}^{\dagger}_{j}\hat{\rho}]$.

The fG Hamiltonian could be rewritten in terms of the correlation function:
\begin{equation}\label{h_c_relation}
    h = -\ln{\frac{1 - C}{C}}.
\end{equation}
which indicates that the eigenvalues of $C$ are bounded between 0 and 1. In the extreme case, when the eigenvalue is 0 (1), the corresponding eigenstate is completely filled (empty) since it should always (never) be occupied according to the energy. 

\subsection{Reshaping Operators}\label{AP:reshaping}

Operators acting on states could be equivalently represented as contraction among states, whereby operators are reshaped into states with suitable auxiliary space. This could be understood in analogue to the thermal field double in general bosonic cases, wherein the bras are reversed to kets making density operators reshaped into a state. Meanwhile, this process of reshaping could also be understood as purification in the context of quantum information \cite{nielsen_chuang_2010}. However, in the fG case, more care should be considered with appending auxiliary space. 

\subsubsection{Bosonic Case}\label{appendix:bosonic_reshaping}
\begin{figure}
    \centering
    \includegraphics[width=1.0\linewidth]{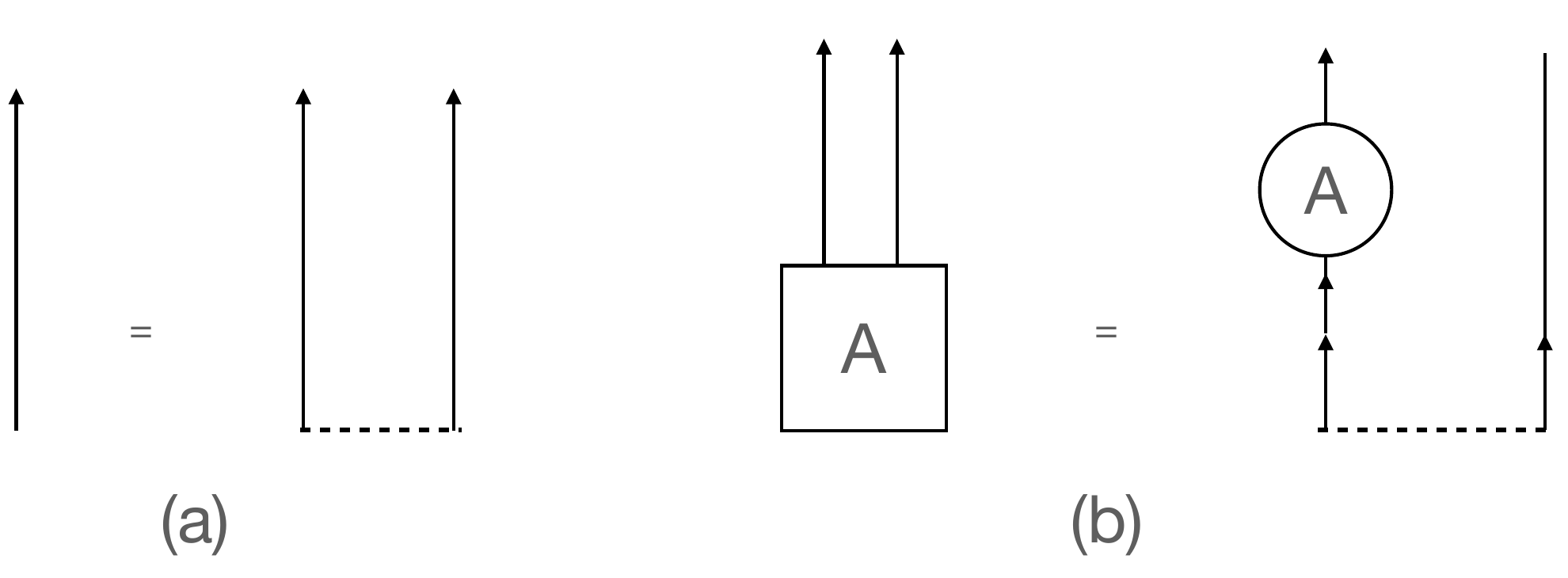}
    \caption{Schematic diagrams for reshaping. (a) reshaping of identity operator. (b) reshaping of an arbitrary operator $A$.}
    \label{fig:purification}
\end{figure}

Reshaping is straightforward in bosonic systems, if there's no further restriction. For instance, we can simply choose the auxiliary space to be identical as the original Hilbert space. To illustrate, we start from purifying an identity operator $\hat{\mathbb{1}}$ to an identity state $\ket{\mathbb{1}}$:
\begin{equation}\label{eq:purify_1}
    \hat{\mathbb{1}} = \sum_{i}\ket{i}\bra{i}\xRightarrow{\text{purified}}\ket{\mathbb{1}} = \sum_{i}\ket{i}\ket{i}_{aux}.
\end{equation}
If the strategy works, the contraction $K(\ket{\mathbb{1}}, \ket{\mathbb{1}'})$ between two identity states $\ket{\mathbb{1}}$ and $\ket{\mathbb{1}'}$ should be an identity state, since $\hat{\mathbb{1}}\ket{\mathbb{1}} = \ket{\mathbb{1}}$. Tracing off the auxiliary space, we obtain the contraction result: 
\begin{align}\label{eq:contract_1}
    K(\ket{\mathbb{1}}, \ket{\mathbb{1}'}) &= \sum_{k}\bra{k}_{aux}\bra{k}_{aux}(\sum_{i,j}\ket{i}\ket{i}_{aux}\ket{j}\ket{j}_{aux}) \notag\\
    & = \sum_{k}\ket{k}\ket{k},
\end{align}
which is as expected. However, the expression of $\ket{\mathbb{1}}$ is not unique, since the unitary transformation of basis $\ket{i}$ will change the expression of $\ket{\mathbb{1}}$. To hold the consistency, we fix the basis in the reshaping of the identity, and derive the reshaping of operators based on the reshaped identity, using the relation for any arbitrary operator $\hat{A}$:
\begin{align}\label{eq:purify_A}
    \ket{A} &= (\hat{A}\otimes\hat{\mathbb{1}})\ket{\mathbb{1}}\notag\\
    &= \sum_{i,j,k,m}A_{ij}(\ket{i}\bra{j}\otimes\ket{k}\bra{k})\ket{m}\ket{m}\notag\\
    &= \sum_{i,m}A_{im}\ket{i}\ket{m}.
\end{align}
The reshaping is demonstrated schematically in Figure \ref{fig:purification}.
The rules of TN diagrams are as following: dashed lines indicate the contraction; bare solid lines represent identity operator; bare lines with arrows are identity states; squares with arrows out are states, while circles with arrows in and out are operators.

\subsubsection{fG case}
Following the same steps in Appendix \ref{appendix:bosonic_reshaping}, we start from the reshaping of a fG identity operator. According to eq. \ref{eq:fG_state}, the identity for a $m$-fermion system is $\hat{\rho}_{\hat{\mathbb{1}}} = \hat{\mathbb{1}}/2^{m}$, which is a maximally entangled state. The corresponding correlation function is $C_{\hat{\mathbb{1}}}=\hat{\mathbb{1}}/2$. In contrast to the bosonic case, simply doubling $C_{\hat{\mathbb{1}}}$ does not give a pure fG state. To satisfy the requirement of a number-conserving fG state, we have to put extra constrains, i.e. the eigenvalues of $C_{\ket{\mathbb{1}}}$ should be 0 or 1. The ways of purification are not unique. The convention we stick to is:
\begin{equation}\label{eq:p_identity}
    C_{\ket{\mathbb{1}}} = \begin{pNiceMatrix}[first-row, first-col]
     & phys & aux\\
     phys & \mathbb{1}/{2} & \mathbb{1}/{2} \\
     aux & \mathbb{1}/{2} & \mathbb{1}/{2}
    \end{pNiceMatrix}\\,
\end{equation}
where the first part of basis corresponds to the physical legs, while the second part belongs to the auxiliary space. Then based on equation \ref{eq:purify_A}, we could reshape any unitary or isometric operator $\hat{U}$ in complex-fermion basis:
\begin{align}\label{eq:purification_C}
    C_{\ket{U}} &= (\hat{U}\oplus\hat{\mathbb{1}})C_{\ket{\mathbb{1}}}(\hat{U}\oplus\hat{\mathbb{1}})^{\dagger}\notag\\
    &= \begin{pmatrix}
        U & \\
         & \mathbb{1}
    \end{pmatrix}\begin{pmatrix}
        \mathbb{1}/{2} & \mathbb{1}/{2} \\
        \mathbb{1}/{2} & \mathbb{1}/{2}
    \end{pmatrix}
    \begin{pmatrix}
            U^{\dagger} & \\
         & \mathbb{1}
    \end{pmatrix}\\
    & = \frac{1}{2}\begin{pmatrix}
        UU^{\dagger} & U\\
        U^{\dagger} & \mathbb{1}
    \end{pmatrix}
\end{align}
The schematics in Figure \ref{fig:purification} also work for the fG case, while all squares representing fG states. The proof is in Appendix \ref{AP:fg_contraction}.

\subsection{Schmidt Decomposition for Pure fG States}\label{ap:schmidtD}

Bipartition of a pure state $\ket{\Psi}$ into two subsystems $A$ and $B$ could be represented as a Schmidt decomposition:
\begin{equation}\label{eq:Schmidt_decomposition_def}
    \ket{\Psi} = \sum_{i}\lambda_{i}\ket{i_{A}}\ket{i_{B}},
\end{equation}
where $\lambda_{i}$ is the real and non-negative Schmidt coefficient, which is valid measure of the entanglement strength. $\ket{i_{A(B)}}$ belongs to orthonormal sets of subsystem A (B). Schmidt decomposition is equivalent to the singular value decomposition (SVD): $\Psi = U_{A}\Sigma V_{B}^{\dagger}$ mathematically, where $U_{A} (V_{B}^{\dagger})$ contains an orthonormal set for subsystem $A$ ($B$). $\Sigma$ is a diagonal matrix with diagonal entries the Schmidt coefficients.

To see the relation between the diagonalization of a submatrix $C_A$ of $C$ in subregion $A$ and Schmidt decomposition, we consider the reduced density matrix $\hat \rho_A = \Tr_B(\ket{\Psi}\bra{
\Psi}) = \sum_i \lambda_i^2 \ket{i_A}\bra{i_A} $, and from the discussion of Gaussian state, $\hat\rho_A = \frac{1}{Z_A} \prod_m e^{- \xi_m \hc_m \hc_m^\dagger}$ where $\xi_m$ are the eigenvalues of restricted correlation matrix in subregion $A$, since the restricted correlation matrix and density matrix are diagonalized simultaneously. Then it is clear the eigenbasis of $C_A$ are just the Schmidt basis of $\hat\rho$ and Schmidt weights $\lambda_i$ can be computed as $\lambda_i^2 = \frac{1}{Z_A}\prod_m e^{-\xi_m\ev{\hc_m\hc_m^\dagger}_{i_A} }$.
 
Given the correlation matrix of a pure fG state $C$, the eigenvalues of $C$ is either 0 or 1, i.e. $U^{\dagger}CU = \begin{pmatrix}
\mathbb{1} & 0 \\
0 & 0
\end{pmatrix}$, where $U$ is the unitary that diagonalizes $C$ with the columns referring to each eigenfunctions respectively. Thus, the correlation matrix could be rewritten using the eigenfunctions $\Phi$ with eigenvalues 1 as $C=\Phi \Phi^{\dagger}$, where those modes in $\Phi$ are referred to as the empty modes in our convention. Dividing elements in $\Phi$ into two parts based on their basis belonging to subsystem $A$ or $B$, the state $\ket{\Phi}$ and correspondingly the correlation $C$ become:
\begin{align}
    \ket{\Phi} &= \begin{pmatrix}
        \Phi_{A}\\
        \Phi_{B}
    \end{pmatrix},\\
    C = \begin{pmatrix}
        C_{AA} & C_{AB} \\
        C_{BA} & C_{BB},
    \end{pmatrix} &= \begin{pmatrix}
        \Phi_{A}\Phi_{A}^{\dagger} & \Phi_{A}\Phi_{B}^{\dagger} \\
        \Phi_{B}\Phi_{A}^{\dagger} & \Phi_{B}\Phi_{B}^{\dagger}
    \end{pmatrix}\label{bi_par_C}.
\end{align}

\begin{figure}
    \centering
    \includegraphics[width=1.0\linewidth]{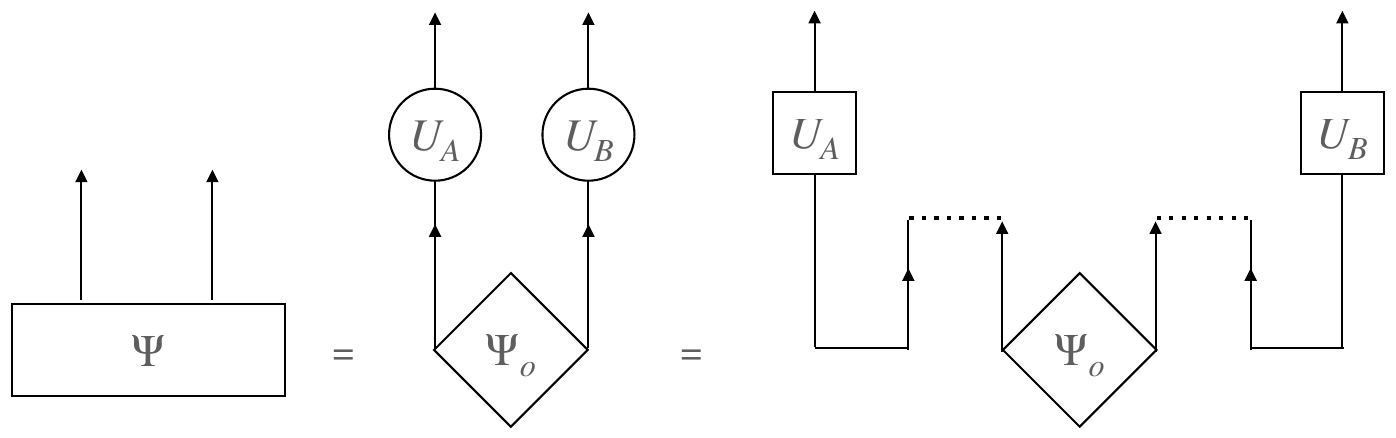}
    \caption{The Schmidt decomposition of a state $\Psi$. The circled $U_{A}$ and $U_{B}$ are isometries correspond to the subsystems $A$ and $B$ after one bipartition. The squared $U_{A}$ and $U_{B}$ are purified states representing the subsystems $A$ and $B$ respectively. The $\Psi_{o}$ contains the Schmidt coefficients of the Schmidt decomposition.}
    \label{fig:schmidt_fG}
\end{figure}
We could further apply SVD to wavefunction $\Phi_{A}$ and $\Phi_{B}$ respectively, where:
\begin{align}\label{SVD_AB}
        \Phi_{A} = U_{A}\Sigma_{A}V_{A}^{\dagger},\\
    \Phi_{B} = U_{B}\Sigma_{B}V_{B}^{\dagger}.
\end{align}
Substituting SVD result into eq. \ref{bi_par_C}, we could obtain the expression for the Schmidt decomposition:
\begin{equation}
    C = \begin{pmatrix}
        U_{A} & \\
         & U_{B}
    \end{pmatrix} \begin{pmatrix}
        \Sigma_{A}\Sigma_{A}^{T} & \Sigma_{A}V_{A}V_{B}^{\dagger}\Sigma_{B}^{T}\\
        \Sigma_{B}V_{B}V_{A}^{\dagger}\Sigma_{A}^{T} & \Sigma_{B}\Sigma_{B}^{T}
    \end{pmatrix}\begin{pmatrix}
        U_{A}^{\dagger} & \\
         & U_{B}^{\dagger}
    \end{pmatrix}.
\end{equation}
However, there exists a phase ambiguity. Although the diagonal entries of the middle matrix are semi-positive and thus fixed, the off-diagonal entries of the middle matrix are not. We claim canonical Schmidt decomposition by requiring the off-diagonal entries non-negative, which could be substantiated by imposing $V_{A} = V_{B}$. Thus, the canonical Schmidt decomposition is:
\begin{equation} \label{eq:diagonal_corr}
    C = \begin{pmatrix}
        U_{A} & \\
         & U_{B}
    \end{pmatrix} \begin{pmatrix}
        \Sigma_{A}\Sigma_{A}^{T} & \Sigma_{A}\Sigma_{B}^{T}\\
        \Sigma_{B}\Sigma_{A}^{T} & \Sigma_{B}\Sigma_{B}^{T}
    \end{pmatrix}\begin{pmatrix}
        U_{A}^{\dagger} & \\
         & U_{B}^{\dagger}
    \end{pmatrix},
\end{equation}
which corresponds to the first equality in Figure \ref{fig:schmidt_fG}.

\subsection{Contraction}\label{AP:fg_contraction}
In this subsection we first present the contraction formula of fG states at the level of correlation matrices, and then show a brief derivation of the formula with Grassmann variables. For a fG state $\hat{\rho}$, which lives in the composite Hilbert space $\mathcal{H}_{c}\otimes\mathcal{H}_{c'}$,  and another fG state in $\mathcal{H}_{c}$,  we could obtain the fG state $\hat{\rho}_{c'}$ belonging to the Hilbert space $\mathcal{H}_{c'}$ by taking contraction between the state $\hat{\rho}$ and another fG state $\hat{\rho}_{c}$ in the space $\mathcal{H}_{c}$. The contraction is equal to $\Tr_{c}[\hat{\rho}_{c}\hat{\rho}]$ up to a normalization factor. This kind of contraction is referred as ``complete contraction'', meaning the all degrees of freedoms in $\hat\rho_c$ are traced out. We further define more general ``incomplete contraction'' by first regarding $\hat\rho_{c''}$ as in a different Hilbert space $\mathcal{H}_{c''}$ and taking tensor product $\hat{\rho}'=\hat\rho\otimes\hat\rho_{c''}\in \mathcal{H}_c\otimes\mathcal{H}_{c'}\otimes\mathcal{H}_{c''}$, and then utilizing complete contraction with a ``contraction kernel'' state $C_{kernel}$ in $\mathcal{H}_{c'}\otimes\mathcal{H}_{c''}$. The contraction kernel is chosen such that the complete contraction of $\hat\rho$ and $\hat\rho_c$ is equal to the incomplete contraction of $\hat\rho$ and particle-hole dual of $\hat\rho_{c''}$. In most cases of this paper we use incomplete contractions, which is consistent as a reverse operation of Schmidt decomposition, i.e., the two states obtained from Schmidt decomposition can be contracted back to the full state.


In the language of correlation matrix, the complete contraction between $\hat{\rho}$ and $\hat{\rho_{c}}$ is given as :
\begin{align}\label{contraction_formula}
    C_{\hat{\rho}_{c'}} &= C_{\hat{\rho}_{00}} - C_{\hat{\rho}_{01}}(C_{\hat{\rho}_{11}} + C_{\hat{\rho}_{c}} - \mathbb{1})^{-1}C_{\hat{\rho}_{10}},\\
    C_{\hat{\rho}} &= \begin{pNiceMatrix}[first-row, first-col]
     & \mathcal{H}_{c'} & \mathcal{H}_{c}\\
     \mathcal{H}_{c'} & C_{\hat{\rho}_{00}} & C_{\hat{\rho}_{01}} \\
     \mathcal{H}_{c} & C_{\hat{\rho}_{10}} & C_{\hat{\rho}_{11}}
    \end{pNiceMatrix}.
\end{align}
With this complete contraction formula, we can do incomplete contractions of two fG states $C_{\hat{\rho}}$ and $C_{\hat{\rho}_{c''}}$ by first taking the direct sum $C_{\hat{\rho}'} = C_{\hat{\rho}}\oplus C_{\hat{\rho}_{c''}}$ and then applying complete contraction of $C_{\hat{\rho}'}$ and  $C_{kernel}$, which is the particle-hole dual of $C_{\ket{\mathbb{1}}}$:
\begin{equation}
 C_{kernel} = C_{\ket{\mathbb{1}}}' = \begin{pmatrix}
        \mathbb{1}/{2} & -\mathbb{1}/{2}\\
        -\mathbb{1}/{2} & \mathbb{1}/{2}
    \end{pmatrix}.
\end{equation}

Next, we could check the equality in Figure \ref{fig:purification} (a), where we employ the complete contraction between $C_{\hat{\rho}'} = C_{\ket{\mathbb{1}}} \oplus C_{\ket{\mathbb{1}}}$ and $C_{kernel}$:

Then we can calculate:
\begin{align} \notag
        & (C_{\hat{\rho}_{11}} + C_{\hat{\rho}_{c}} - \mathbb{1})^{-1} \\ \notag
        & = (\frac{1}{2}\begin{pmatrix}
            \mathbb{1} & 0\\
            0 & \mathbb{1}
        \end{pmatrix} + \frac{1}{2}\begin{pmatrix}
            \mathbb{1} & -\mathbb{1}\\
            -\mathbb{1} & \mathbb{1}
        \end{pmatrix} - \begin{pmatrix}
            \mathbb{1} & 0 \\
            0 & \mathbb{1}
        \end{pmatrix})^{-1}\\ 
        & = -2\begin{pmatrix}
            0 & \mathbb{1}\\
            \mathbb{1} & 0
        \end{pmatrix},\\ \notag
        & C_{\hat{\rho}_{c'}}\\ \notag
        & = \frac{1}{2}\begin{pmatrix}
            \mathbb{1} & 0\\
            0 & \mathbb{1}
        \end{pmatrix}  - \frac{1}{2}\begin{pmatrix}
            \mathbb{1} & 0\\
            0 & \mathbb{1}
        \end{pmatrix}(-2\begin{pmatrix}
            0 & \mathbb{1}\\
            \mathbb{1} & 0
        \end{pmatrix})\frac{1}{2}\begin{pmatrix}
            \mathbb{1} & 0\\
            0 & \mathbb{1}
        \end{pmatrix} \\ 
        & = \frac{1}{2}\begin{pmatrix}
            \mathbb{1} & \mathbb{1}\\
            \mathbb{1} & \mathbb{1}
        \end{pmatrix}.
\end{align}
Similarly, we could also verify the contraction in Figure \ref{fig:schmidt_fG}, where we apply the complete contraction of $C_{\hat{\rho}'} = C_{\ket{U_{A}}} \oplus C_{\ket{\Psi_{o}}} \oplus C_{\ket{U_{B}}}$ and $C_{kernel}$:
\begin{align} \notag
    & (C_{\hat{\rho}_{11}} + C_{\hat{\rho}_{c}} - \mathbb{1})^{-1} \\ \notag
    & = (\begin{pmatrix}
         \Sigma_{A}\Sigma_{A}^{T} & \Sigma_{A}\Sigma_{B}^{T} & &\\
         \Sigma_{B}\Sigma_{A}^{T} & \Sigma_{B}\Sigma_{B}^{T} & &\\
         & & \frac{1}{2}\mathbb{1} & \\
         & & & \frac{1}{2}\mathbb{1}
    \end{pmatrix} + \\ \notag
    &\frac{1}{2}\begin{pmatrix}
                    \mathbb{1} & & -\mathbb{1} & \\
             & \mathbb{1} & & -\mathbb{1}\\
             -\mathbb{1} & & \mathbb{1} & \\
             & -\mathbb{1} & & \mathbb{1}
    \end{pmatrix} - \begin{pmatrix}
        \mathbb{1} & & &\\
         & \mathbb{1} & & \\
         & & \mathbb{1} & \\
         & & & \mathbb{1}
    \end{pmatrix})^{-1}\\ 
    & = -2\begin{pmatrix}
        & & \mathbb{1} & \\
        & & & \mathbb{1} \\
        \mathbb{1} &  & 2\Sigma_{A}\Sigma_{A}^{T}-\mathbb{1} & 2\Sigma_{A}\Sigma_{B}^{T}\\
        & \mathbb{1} & 2\Sigma_{B}\Sigma_{A}^{T} & 2\Sigma_{B}\Sigma_{B}^{T} - \mathbb{1}
    \end{pmatrix}\\
    & C_{\hat{\rho}_{c'}}\\ \notag
    & = \frac{1}{2}\begin{pmatrix}
        U_{A}U_{A}^{\dagger} & \\
        & U_{B}U_{B}^{\dagger}
    \end{pmatrix} + \\ \notag
    &\frac{1}{2}\begin{pmatrix}
        U_{A} & \\
        & U_{B}
    \end{pmatrix}2(2\begin{pmatrix}
        \Sigma_{A}\Sigma_{A}^{T} & \Sigma_{A}\Sigma_{B}^{T}\\
        \Sigma_{B}\Sigma_{A}^{T} & \Sigma_{B}\Sigma_{B}^{T}
    \end{pmatrix} - \mathbb{1})\frac{1}{2}\begin{pmatrix}
        U_{A}^{\dagger} & \\
        & U_{B}^{\dagger}
    \end{pmatrix} \\ 
    & = \begin{pmatrix}
        U_{A} & \\
         & U_{B}
    \end{pmatrix} \begin{pmatrix}
        \Sigma_{A}\Sigma_{A}^{T} & \Sigma_{A}\Sigma_{B}^{T}\\
        \Sigma_{B}\Sigma_{A}^{T} & \Sigma_{B}\Sigma_{B}^{T}
    \end{pmatrix}\begin{pmatrix}
        U_{A}^{\dagger} & \\
         & U_{B}^{\dagger}
    \end{pmatrix},
\end{align}
where the final result is exactly the expression in equation \ref{eq:diagonal_corr}.

Now we present a derivation of the formula with Grassmann integrations, based on the Grassmann representation of partial trace in \cite{Bravyi2004}.  Consider a density matrix $\hat\rho$ of $m$ fermionic degrees of freedoms, the Grassmann representation is 
\begin{equation}
    w_\rho(\eta) = e^{-\frac{i}{2}\eta^T M \eta},\quad \eta^T = (\eta_0, ..., \eta_{2m-1}),
\end{equation}
where $\eta$ is a column vector of real Grassmann numbers, and $M_{ij} = -\frac{i}{2}\ev{[\hat\gamma_i,\hat\gamma_j]}$ is  usually called covariance matrix with Majorana operators defined through $\hc_k = \frac{1}{2}(\hat \gamma_{2k} + i\hat\gamma_{2k+1})$. By doing a bit algebra we can turn to a complex Grassmann representaion, such that 
\begin{equation}
    w_\rho(\xi) = e^{- \bar\xi^T (2C-1) \xi},
\end{equation}
where $\bar\xi$ is the complex conjugate of $\xi$. For two states $\rho, \rho_c$ as defined before, the complex Grassmann representation are:
\begin{equation}
\begin{split}
        w_\rho(\xi) &= e^{-\bar\xi^T (2C_\rho-1) \xi},\quad \xi^T = (\xi_c, \xi_{c'}) \\
        w_{\rho_c}(\zeta) &= e^{-\bar\zeta^T (2C_{\rho_c}-1) \zeta},
\end{split}
\end{equation}
Then according to \cite{Bravyi2004}, the partial trace $\Tr_{c}[\hat\rho\hat\rho_c]$ is represented by
\begin{equation}
    \Tr_{c}[\hat\rho\hat\rho_c](\xi_{c'}) = \int D\xi_c D\bar\xi_c D\zeta D\bar\zeta \quad e^{\bar\xi_c^T \zeta - \bar\zeta^T\xi_c} w_\rho(\xi) w_{\rho_c}(\zeta).
\end{equation}
Putting the form of $C_\rho$ and $C_{\rho_c}$ together, the right hand side turns out to be the following Gaussian integration:
\begin{equation}
\begin{split}
    &\int D\xi_c D\bar\xi_c D\zeta D\bar\zeta \\
    \times & e^{ 
    -(\bar\xi_{c'}^T \  \bar\xi_c^T \  \bar\zeta^T) \left(
 \begin{array}{ccc}
 2C_{00} - 1 & 2C_{01} & 0 \\
 2C_{10} & 2C_{11}- 1  & -1 \\
 0 & 1 & 2C_{\rho_c}-1 \\
 \end{array}\right) \left(
 \begin{aligned}
 & \xi_{c'} \\
 & \xi_c \\
 & \zeta \\
 \end{aligned}
 \right)} .
\end{split}
\end{equation}
Using Gaussian integration formula, we obtain a Gaussian state which is represented by 
\begin{equation}
    2C_{c'}-1 = 2C_{00} - 1 - 2C_{01} \left(  2C_{11}- 1 +(2C_{\rho_c}-1)^{-1}
	\right)^{-1} 2C_{01}^\dagger
\end{equation}
For the case that $\rho_c$ is pure, we can further simplify this result by noticing that $(2C_{\rho_c}-1)^2 = 1$, so it becomes the contraction formula eq. (\ref{contraction_formula}).

\subsection{Deformation}\label{AP:deformation}
In general the correlation matrix $C = \ev{\hc_i\hc_j^\dagger}$ of a fermionic state has eigenvalues between 0 and 1. In particular, if the state is pure, then the eigenvalues of the correlation matrix have to be 0 or 1 if and only if the state is Gaussian. In the main text, we met a question of stacking the local tensor of Wannier states to get a local tensor of the whole ground state, which physically means taking a product of these Wannier states. At the level of correlation matrices, it corresponds to the sum of each individual correlation matrices, each regarded as being embedded in a local Hilbert space of all physical and bond degrees of freedom. But the literal sum of two correlation matrices may not be a valid correlation matrix for a physical state, i.e., the eigenvalues of the sum could be negative. To obtain a physically valid correlation matrix of a pure free-fermion state, we need a {\itshape deformation} to make sure all eigenvalues that are not equal to 1 to be exactly equal to 0.

The reason why the sum of two correlation matrices is not a valid correlation matrix is the possible non-orthogonality of the two states to be summed. To see this, let $C_1$ and $C_2$ be two correlation matrices of pure fG states. The fermionic degrees of freedom are $\hc_1^\dagger, ..., \hc_k^\dagger$, $k\in \mathbb{N^+}$. Then the eigen-modes of $C_1$ could be represented by some normalized row vectors $\Psi_1$, such that 
\begin{equation}\label{eq:corrcalc}
C_1 = \mathbb{1} - \Psi_1^\dagger\Psi_1,
\end{equation}
 
\begin{equation}
    \Psi_1 = \left(\begin{array}{cccc}
    m_{1, 1} & m_{1, 2} &\cdots & m_{1, k} \\
    m_{2, 1} & m_{2, 2} &\cdots & m_{2, k} \\
    \vdots   & \vdots & \vdots & \vdots \\
    m_{r_1, 1} & m_{r, 2} &\cdots & m_{r_1, k} \\
    \end{array}
    \right), k \ge r,
\end{equation}
where $r_1$ is the rank of $C_1$. Each row vector in $\Psi_1$ denotes an eigen-mode $\hat d_i^\dagger = \sum_{j}m_{i,j} \hc_j^\dagger$, so the Dirac symbol of the corresponding product state $\ket{\Psi_1} = \hat d_1^\dagger \cdots \hat d_r^\dagger \ket{0}$ has coefficients being the Slater determinants of any $r$-column submatrix of $\Psi_1$, or say, the minors of order $r$. Similar one can obtain row vectors $\Psi_2$ of rank $r_2$ for the correlation matrix $C_2$ . We see that each time we add a row vector to $\Psi_1$, we are adding an occupation of a mode to the state $\ket{\Psi_1}$. However, for the relation $C = \mathbb{1} - \Psi^\dagger\Psi$ to work, there is a necessary condition: all normalized row vectors should be orthogonal to each other (we assume that all row vectors are already linearly independent).  For instance, given two modes $\tilde d_1^\dagger = a_1 \hc_1^\dagger + a_2 \hc_2^\dagger, \tilde d_2^\dagger = b_1 \hc_1^\dagger + b_3 \hc_3^\dagger$, the product state is $
\ket{\tilde d_1^\dagger \tilde d_2^\dagger} = \mathcal{N}^{-1} (-a_2b_1 \ket{110}+ a_1b_3\ket{101} + a_2b_3\ket{011})$,with normalization factor $\mathcal{N} = \sqrt{a_2^2b_1^2 + a_2^2b_3^2 +a_1^2b_3^2}$ (we assume these coefficients to be real and $a_1^2 + a_2^2 =1, b_1^2+b_3^2 = 1$). It is clear to get the corresponding correlation matrix $C$, which is
\begin{equation}
   C = \mathbb{1} - \frac{1}{1-a_1^2b_1^2}\left( \begin{array}{ccc}
        a_1^2+b_1^2-2a_1^2b_1^2 & a_1a_2b_3^2 & -b_1b_3a_2^2\\
         a_1a_2b_3^2 & 1-a_1^2 & a_1a_2b_1b_3 \\
         -b_1b_3a_2^2 & a_1a_2b_1b_3 & 1-b_1^2 \\
    \end{array}\right), 
\end{equation}
while if we put the naive row vectors 
\begin{equation}
    \left(\begin{array}{ccc}
    a_1 & a_2 & \\
    b_1 & & b_3 \\
    \end{array}
    \right)
\end{equation}
into eq. (\ref{eq:corrcalc}) it will give 
\[ \tilde C = 
\mathbb{1} - \left(\begin{array}{ccc}
 a_1^2+b_1^2 & a_1a_2 & b_1b_3 \\
 a_1a_2 & a_2^2 & 0 \\
 b_1b_3 & 0 & b_3^2 \\
\end{array}\right)
\]
which does not work. Instead, the proper row vectors should be the ones after doing the Gram-Schmidt process and normalization, 
\begin{equation}
    \Psi = \left(
    \begin{array}{ccc}
    a_1 & a_2 & 0 \\
    \frac{a_2^2 b_1}{\sqrt{1-a_1^2b_1^2}} & \frac{-a_1a_2b_1}{\sqrt{1-a_1^2b_1^2}} & \frac{b_3}{\sqrt{1-a_1^2b_1^2}} \\
    \end{array}
    \right).
\end{equation}

In general, consider a set of non-orthonormalized modes $\{\tilde d_1^\dagger, \tilde d_2^\dagger, ..., \tilde d_r^\dagger\}$, from which we get the non-orthonormalized row vectors $\tilde \Psi$, the corresponding orthonormalized row vectors $\Psi$ will give the correct correlation matrix of the product state $\prod_{i=1}^{r} \tilde d_i^\dagger \ket{0}$. To see that, we make use of the QR decomposition of the transpose matrix $ \tilde\Psi^T_{k\times r} = Q_{k\times k}R_{k\times r}$ where $Q$ is unitary and $R$ is upper triangular in the first $r$ rows and zero in the remaining $k-r$ rows. So the state $\prod_i \tilde d_i^\dagger \ket{0}= \prod_{i}\sum_l \tilde m_{i,l}c_l^\dagger \ket{0}$ represented by $\Psi$ is equivalent to 

$$\prod_i \sum_{j\le i,l} (R^T)_{i,j}(Q^T)_{j,l} \hc_l^\dagger\ket{0} =\prod_{i} \sum_{j\le i} R_{j,i}\hat d_j^\dagger \ket{0}\propto \prod_{i}\hat d_i^\dagger\ket{0}.$$
Since $Q$ is unitary, $\hat d_j^\dagger \equiv Q_{l,j}\hc_l^\dagger$ is a canonical basis, thus this state is equivalent to $\prod_{i}\hat d_i^\dagger$ up to a normalization. We also see from the QR decomposition that the if the row vectors $ \Psi$ are not orthonormal, the $\Psi^\dagger \Psi = Q^* R^* R^T Q^T $ will have the same non-vanishing eigenvalues as $R^*R^T$ which are only guaranteed to be positive but not necessarily equal to identity.

Therefore, we see the naive literal ``sum'' of $C_1$ and $C_2$ will not work since in general the eigen-modes of $C_1$ and those of $C_2$ are not orthogonal. Suppose the joint row vectors $$ \tilde \Psi = \left( \begin{array}{c}
        \Psi_1 \\
        \Psi_2
\end{array}\right)$$ has rank $r_1+r_2 < k$, then we can find the eigenbasis of $\tilde C = \mathbb{1} - \tilde\Psi^\dagger  \tilde\Psi$, that is, $\tilde C = U^\dagger\Sigma_{\tilde C} U$ with eigenvalues in $\Sigma_{\tilde C} = \text{diag}[\lambda_1, ..., \lambda_{r_1+r_2}, 1,...,1]$. Notice that the eigenvectors corresponding to eigenvalue 1 will be the empty modes, which will not be changed under the Gram-Schmidt process. So taking the eigenvectors except those corresponding to the eigenvalue 1 as the $\Psi$, we can obtain the a valid correlation matrix representing the product state $\ket{\Psi_1}\ket{\Psi_2}$. This is equivalent to deforming all the non-unity eigenvalues of $\tilde C$ to zero.  

Numerically one may worry that the exact unity may be not easy to distinguish, if some occupied modes of non-orthogonal $\tilde \Psi$ make some eigenvalues of $\tilde C = \mathbb{1} - \tilde\Psi^\dagger \tilde\Psi $ close to identity numerically. Luckily this situation will not happen in the main text stacking the local tensors from Wannier functions. Physically the eigenvalues of $\Psi^\dagger\Psi$ will be close to zero only when some row vectors in $\Psi$ are close to be collinear, i.e., $\vec{m}_1\approx \lambda \vec{m}_2, \lambda\ne 0 $. A Wannier function of the state is not translational invariant, so local tensors at different locations are not possibly collinear. In fact, the spectrum of eigenvalues in the stacked tensors turn out to exhibit clear gaps separating the unity and non-unity values. Thus there is no numerical ambiguity in the deformation process.

\end{document}